# Succinct Representation of Codes with Applications to Testing

Elena Grigorescu*    Tali Kaufman†    Madhu Sudan‡

October 14, 2018

## Abstract

Motivated by questions in property testing, we search for linear error-correcting codes that have the "single local orbit" property: i.e., they are specified by a single local constraint and its translations under the symmetry group of the code. We show that the dual of every "sparse" binary code whose coordinates are indexed by elements of $\mathbb{F}_{2^n}$ for prime $n$, and whose symmetry group includes the group of non-singular affine transformations of $\mathbb{F}_{2^n}$, has the single local orbit property. (A code is said to be *sparse* if it contains polynomially many codewords in its block length.) In particular this class includes the dual-BCH codes for whose duals (i.e., for BCH codes) simple bases were not known. Our result gives the first short ($O(n)$-bit, as opposed to the natural $\exp(n)$-bit) description of a low-weight basis for BCH codes.

The interest in the "single local orbit" property comes from the recent result of Kaufman and Sudan (STOC 2008) that shows that the duals of codes that have the single local orbit property under the affine symmetry group are locally testable. When combined with our main result, this shows that all sparse affine-invariant codes over the coordinates $\mathbb{F}_{2^n}$ for prime $n$ are locally testable.

If, in addition to $n$ being prime, if $2^n - 1$ is also prime (i.e., $2^n - 1$ is a Mersenne prime), then we get that every sparse *cyclic* code also has the single local orbit. In particular this implies that BCH codes of Mersenne prime length are generated by a single low-weight codeword and its cyclic shifts.

In retrospect, the single local orbit property has been central to most previous results in algebraic property testing. However, in the previous cases, the single local property was almost "evident" for the code in question (the single local constraint was explicitly known, and it is a simple algebraic exercise to show that its translations under the symmetry group completely characterize the code). Our work gives an alternate proof of the single local orbit property, effectively by counting, and its effectiveness is demonstrated by the fact that we are able to analyze it in cases where even the local constraint is not "explicitly" known. Our techniques involve the use of recent results from additive number theory to prove that the codes we consider, and related codes emerging from our proofs, have high distance. We then combine these with the MacWilliams identities and some careful analysis of the invariance properties to derive our results.

*MIT CSAIL, `elena_g@mit.edu`.
†MIT CSAIL, `kaufmant@mit.edu`.
‡MIT CSAIL, `madhu@mit.edu`.

# 1 Introduction

Motivated by questions about the local testability of some well-known error-correcting codes, in this paper we examine their "invariance" properties. Invariances of codes are a well-studied concept (see, for instance, [16, Chapters 7, 8.5, and 13.9]) and yet we reveal some new properties of BCH codes. In the process we also find broad classes of sparse codes that are locally testable. We describe our problems and results in detail below.

A code $C \subseteq \mathbb{F}_2^N$ is said to be locally testable if membership of a word $w \in \mathbb{F}_2^N$ in the code $C$ can be checked probabilitistically by a few probes into $w$. The famed "linearity test" of Blum, Luby and Rubinfeld [2] may be considered the first result to show that some code is locally testable. Locally testable codes were formally defined by Rubinfeld and Sudan [17]. The first substantial study of locally testable codes was conducted by Goldreich and Sudan [9], where the principal focus was the construction of locally testable codes of high rate. Local testing of codes is effectively equivalent to property testing [17, 8] with the difference being that the emphasis here is when $C$ is an error-correcting code, i.e., elements of $C$ are pairwise far from each other.

A wide variety of "classical" codes are by now known to be locally testable, including Hadamard codes [2], Reed-Muller codes of various parameters [17, 1, 13, 10], dual-BCH codes [11, 14], turning attention to the question: What broad characteristics of codes are necessary, or sufficient, for codes to be locally testable. One characteristic explored in the recent work of Kaufman and Sudan [15] is the "invariance group" of the code, which we describe next.

Let $[N]$ denote the set of integers $\{1, \ldots, N\}$. A code $C \subseteq \mathbb{F}_2^N$ is said to be invariant under a permutation $\pi : [N] \to [N]$ if for every $a = \langle a_1, \ldots, a_N \rangle \in C$, it is the case that $a \circ \pi = \langle a_{\pi(1)}, \ldots, a_{\pi(N)} \rangle$ is also in $C$. The set of permutations under which any code $C$ is invariant forms a group under composition and we refer to it as the invariant group. [15] suggested that the invariant group of a code may play an important role in its testability. They supported their suggestion by showing that if the invariant group is an "affine group", then a "linear" code whose "dual" has the "single local orbit" property is locally testable. We explain these terms (in a restricted setting) below.

Let $N = 2^n$ and let $C \subseteq \mathbb{F}_2^N$ be a code. In this case we can associate the coordinate set $[N]$ of the code $C$ with the field $\mathbb{F}_{2^n}$. Now consider the permutations $\pi : \mathbb{F}_{2^n} \to \mathbb{F}_{2^n}$ of the form $\pi(x) = \alpha x + \beta$ where $\alpha \in \mathbb{F}_{2^n} - \{0\}$ and $\beta \in \mathbb{F}_{2^n}$. This set is closed under composition and we refer to this as the *affine group*. If $C$ is invariant under every $\pi$ in the affine group, then we say that $C$ is *affine-invariant*. We say that $C$ is *linear* if it is a vector subspace of $\mathbb{F}_2^N$. The *dual* of $C$, denoted $C^\perp$, is the null space of $C$ as a vector space.

We now define the final term above, namely, the "single local orbit property". Let $G$ be a group of permutations mapping $[N]$ to $[N]$. For $b \in \mathbb{F}_2^N$, let its *weight*, denoted wt($b$), be the number of non-zero elements of $b$. A code $C$ is said to have the *$k$-single orbit property* under $G$ if there exists an element $b \in \mathbb{F}_2^N$ of weight at most $k$ such that $C = \text{Span}(\{b \circ \pi | \pi \in G\})$, where $\text{Span}(S) = \{\sum_i c_i b_i | c_i \in \mathbb{F}_2, b_i \in S\}$. Two groups are of special interest to us in this work. The first is the affine group on $\mathbb{F}_{2^n}$. A second group of interest to us is the "cyclic group" on $\mathbb{F}_{2^n}^* = \mathbb{F}_{2^n} - \{0\}$ given by the permutations $\pi_a(x) = ax$ for $a \in \mathbb{F}_{2^n}^*$. (Note that if $\omega$ is a multiplicative generator of $\mathbb{F}_{2^n}^*$ and the coordinates of $C$ are ordered $\langle \omega, \omega^2, \ldots, \omega^{2^n-1} = 1 \rangle$ then each $\pi_a$ is simply a cyclic permutation.)

The invariance groups of codes are well-studied objects. In particular codes that are invariant under cyclic permutations, known as cyclic codes, are widely studied and include many common algebraic codes (under appropriate ordering of the coordinates and with some slight modifications, see [18] or [16]). The fact that many codes are also affine-invariant is also explicitly noted and used in the literature [16].

Conditions under which codes have the single-orbit property under any given group, seem to be less well-studied. This is somewhat surprising given that the single-orbit property implies very succinct (nearly explicit) descriptions (of size $k \log N$ as opposed to $N^2$) of bases for codes (that have the $k$-single orbit property under some standard group). Even for such commonly studied codes such as the BCH codes such explicit descriptions of bases were not known prior to this work. In retrospect, the single orbit property was being exploited in previous results in algebraic property testing [2, 17, 1, 13, 10] though this fact was not explicit until the work of [15].

In this work we explore the single orbit property under the affine group for codes on the coordinate set $\mathbb{F}_{2^n}$, as also the single orbit property under the cyclic group for codes over $\mathbb{F}_{2^n}^*$. We show that the dual of every "sparse" affine-invariant code (i.e., codes with at most polynomially many codewords in $N$) has the $k$-single orbit property under the affine group for some constant $k$, provided $N = 2^n$ for prime $n$ (see Theorem 4). When $N - 1$ is also prime, it turns out that the duals of sparse codes have the $k$-single orbit property under the cyclic group for some constant $k$ yielding an even stronger condition on the basis (see Theorem 5). Both theorems shed new light on well-studied codes including BCH codes.

In particular the first theorem has immediate implications for testing and shows that every sparse affine invariant code is locally testable. This merits comparison with the results of [14] who show that sparse high-distance codes are locally testable. While syntactically the results seem orthogonal (ours require affine-invariance whereas theirs required high-distance) it turns out (as we show in this paper) that all the codes we consider do have high-distance. Yet for the codes we consider our results are more constructive in that they not only prove the "existence" of a local test, but give a much more "explicit" description of the tester: Our tester is described by a single low-weight word in the dual and tests that a random affine permutation of this word is orthogonal to the word being tested. [1]

Given a code of interest to us, we first study the algebraic structure of the given code by representing codewords as polynomials and studying the degree patterns among the support of these polynomials. We interpret the single orbit property in this language; and this focusses our attention on a collection of closely related codes. We then turn to recent results from additive number theory [4, 3, 6, 5, 7] and apply them to the dual of the given code, as well as the other related codes that arise from our algebraic study, to lower bound their distance. In turn, using the MacWilliams identities (as in prior work [14]) this translates to some information on the weight-distribution of the given code and the related ones. Some simple counting now yields that the given code must have the single-orbit property.

We believe that our techniques are of interest, beyond just the theorems they yield. In particular we feel that techniques to assert the single-orbit property are quite limited in the literature. Indeed in all previous results [2, 17, 1, 13, 10] this property was "evident" for the code: The local constraint whose orbit generated a basis for all constraints was explicitly known, and the algebra needed to prove this fact was simple. Our results are the first to consider the setting where the basis is not explicitly known (even after our work) and manages to bring in non-algebraic tools to handle such cases. We believe that the approach is potentially interesting in broader settings.

---

[1]In contrast the tester of [14] was less "explicit". It merely proved the existence of many low weight codewords in the dual of the code being tested and proved that the test which picked one of these low-weight codewords uniformly at random and tested orthogonality of the given word to this dual codeword was a sound test.

## 2 Definitions and main results

We recall some basic notation. $[N]$ denotes the set $\{1, \ldots, N\}$. $\mathbb{F}_q$ denotes the finite field with $q$ elements and $\mathbb{F}_q^*$ will denote the non-zero elements of this field. We will consider codes contained in the vector space $\mathbb{F}_2^N$. For a word $a = \langle a_1, \ldots, a_N \rangle \in \mathbb{F}_2^N$ its support is the set $\mathrm{Supp}(a) = \{i | a_i \neq 0\}$ and its weight is the quantity $\mathrm{wt}(a) = |\mathrm{Supp}(a)|$. For $a = \langle a_i \rangle_i$, and $b = \langle b_i \rangle_i \in \mathbb{F}_2^N$ define the *relative distance* between $a, b$ as $\delta(a,b) = \frac{1}{N} |\{i \mid a_i \neq b_i\}|$. Note $\delta(a,b) = \frac{\mathrm{wt}(a-b)}{N}$.

A binary code $\mathcal{C}$ is a subset of $\mathbb{F}_2^N$. The (relative) distance of $\mathcal{C}$ is $\delta(\mathcal{C}) = \min_{a,b \in \mathcal{C}; a \neq b} \{\delta(a,b)\}$.

For a set of vectors $S = \{v_1, \ldots, v_k\} \subseteq \mathbb{F}_2^N$, let $\mathrm{Span}(S) = \{\sum_{i=1}^k \alpha_i v_i | \alpha_1, \ldots, \alpha_k \in \mathbb{F}_2\}$ denote the linear span of $S$. $\mathcal{C}$ is a *linear* code if its codewords form a vector space in $\{0,1\}^N$ over $\mathbb{F}_2$, i.e., if $\mathrm{Span}(\mathcal{C}) = \mathcal{C}$. For $a, b \in \mathbb{F}_2^N$, let $a \cdot b = \sum_i a_i b_i$ denote the inner product of $a$ and $b$. The *dual* of $\mathcal{C}$ is the code $\mathcal{C}^\perp = \{b \in \mathbb{F}_2^N \mid b \cdot a = 0, \ \forall a \in \mathcal{C}\}$.

We will alternate between viewing $a \in \mathbb{F}_2^N$ as a vector $a = \langle a_1, \ldots, a_N \rangle$ and as a function $a : D \to \mathbb{F}_2$ where $D$ will be some appropriate domain of size $N$. Two particular domains of interest to us will be $\mathbb{F}_{2^n}$ and $\mathbb{F}_{2^n}^*$.

### 2.1 Invariance and the single local orbit property

Let $a \in \mathbb{F}_2^N$ be viewed as a function $a : D \to \mathbb{F}_2$ for some domain $D$ of size $N$. Let $\pi : D \to D$ be a permutation of $D$. The $\pi$-rotation of $a$ is the function $a \circ \pi : D \to \mathbb{F}_2$ given by $a \circ \pi(i) = a(\pi(i))$ for every $i \in D$.

Let $D$ be a set of size $N$ and let $\mathbb{F}_2^N$ denote the set of functions from $D \to \mathbb{F}_2$. A code $\mathcal{C} \subseteq \mathbb{F}_2^N$ is said to be *invariant* under a permutation $\pi : D \to D$ if for every $a \in \mathcal{C}$, it is the case that $a \circ \pi \in \mathcal{C}$. The set of permutations under which a code $\mathcal{C}$ is invariant forms a group under composition and we refer to it as the invariant group of a code.

We will be interested in studying codes that are invariant under some well-studied groups (i.e., whose invariant groups contain some well-studied groups). Two groups of interest to us are the affine group over $\mathbb{F}_{2^n}$ and the cyclic group over $\mathbb{F}_{2^n}^*$. In what follows we let $N = 2^n$ and view $\mathbb{F}_2^N$ as the set of functions from $\mathbb{F}_{2^n}$ to $\mathbb{F}_2$ and $\mathbb{F}_2^{N-1}$ as the set of functions from $\mathbb{F}_{2^n}^*$ to $\mathbb{F}_2$.

**Definition 1 (Affine invariance)** *A function $\pi : \mathbb{F}_{2^n} \to \mathbb{F}_{2^n}$ is an* affine permutation *if there exist $\alpha \in \mathbb{F}_{2^n}^*$ and $\beta \in \mathbb{F}_{2^n}$ such that $\pi(x) = \alpha x + b$. The* affine group *over $\mathbb{F}_{2^n}$ consists of all the affine permutations over $\mathbb{F}_{2^n}$. A code $\mathcal{C} \subseteq \mathbb{F}_2^N$ is said to be* affine invariant *if the invariant group of $\mathcal{C}$ contains the affine group.*

**Definition 2 (Cyclic invariance)** *A function $\pi : \mathbb{F}_{2^n}^* \to \mathbb{F}_{2^n}^*$ is a* cyclic permutation *if it is of the form $\pi(x) = \alpha x$ for $\alpha \in \mathbb{F}_{2^n}^*$.* [2] *The* cyclic group *over $\mathbb{F}_{2^n}^*$ consists of all the cyclic permutations over $\mathbb{F}_{2^n}^*$. A code $\mathcal{C} \subseteq \mathbb{F}_2^{N-1}$ is said to be* cyclic invariant *(or simply cyclic) if the invariant group of $\mathcal{C}$ contains the cyclic group.*

Many well-known families of codes (with minor variations) are known to be affine-invariant and/or cyclic. In particular BCH codes are cyclic and Reed-Muller codes are affine-invariant. Furthermore under a simple "extension" operation BCH codes become affine-invariant, and vice versa under a simple puncturing operation, Reed-Muller codes become cyclic. We elaborate on these later.

In this paper our aim is to show that certain families of affine-invariant and cyclic codes have a simple description, that we call a "single-orbit description". We define this term next.

**Definition 3 ($k$-single orbit code)** *Let $\mathbb{F}_2^N$ be the collection of functions from $D$ to $\mathbb{F}_2$ for some domain $D$. Let $G$ be a group of permutations from $D$ to $D$. A linear code $\mathcal{C} \subseteq \mathbb{F}_2^N$ is said to have the $k$-single orbit property under the group $G$ if there exists $a \in \mathcal{C}$ with $\mathrm{wt}(a) \leq k$ such that $\mathcal{C} = \mathrm{Span}(\{a \circ \pi | \pi \in G\})$.*

In particular the $k$-single orbit property under the affine group has implications to testing that we discuss in Section 2.3.

## 2.2 Main results

Our main results show that, under certain conditions, duals of "sparse" codes have the single orbit property for small $k$. By "sparse" we mean that the code has only polynomially many codewords in the length of the codewords.

Our first result considers affine-invariant codes.

**Theorem 4 (Single orbit property in affine-invariant codes)** *For every $t > 0$ there exists a $k = k(t)$ such that for every prime $n$ the following holds: Let $N = 2^n$ and $\mathcal{C} \subseteq \mathbb{F}_2^N$ be a linear affine-invariant code containing at most $N^t$ codewords. Then $\mathcal{C}^\perp$ has the $k$-single orbit property under the affine group.*

Next we present our main theorem for cyclic codes.

**Theorem 5 (Single orbit property in cyclic codes)** *For every $t$ there exists a $k$ such that the following holds: Let $n$ be such that $2^n - 1$ is prime. Let $\mathcal{C} \subseteq \mathbb{F}_2^{N-1}$ be a linear, cyclic invariant, code with at most $N^t$ codewords. Then $\mathcal{C}^\perp$ has the $k$-single orbit property under the cyclic group.*

We remark that it is not known if there are infinitely many $n$ such that $2^n - 1$ is prime. Of course if there are only finitely many such primes then our theorem becomes "trivial". Nevertheless, as things stand, the question of whether the number of such primes is infinite or not is unresolved (and indeed there are conjectures suggesting there are infinitely many such primes), and so unconditional result should remain interesting.

---

[2] Note that this is a permutation of $\mathbb{F}_{2^n}^*$ if the elements of $\mathbb{F}_{2^n}^*$ are enumerated as $\langle \omega, \omega^2, \ldots, \omega^{N-1} \rangle$ where $\omega$ is a multiplicative generator of $\mathbb{F}_{2^n}^*$.

## 2.3 Implications to property testing

It follows from the work of [15] that codes with a single local orbit under the affine symmetry group are locally testable. We recall some basic definitions below and summarize the implication of our main theorem to testability.

**Definition 6 (Locally testable code [9])** *A code $\mathcal{C} \subseteq \mathbb{F}_2^N$ is $(k, \alpha)$-locally testable if there exists a probabilistic algorithm $T$ called the tester that, given oracle access to a vector $v \in \mathbb{F}_2^N$ makes at most $k$, queries to the oracle for $v$ and accepts $v \in \mathcal{C}$ with probability 1, while rejecting $v \notin \mathcal{C}$ with probability at least $\alpha \cdot \delta(v, \mathcal{C})$. $\mathcal{C}$ is said to be locally testable if there exist $k < \infty$ and $\alpha > 0$ such that $\mathcal{C}$ is $(k, \alpha)$-locally testable.*

We note that the above definition corresponds to the strong definition of local testability ([9, Definition 2.2]). We now state the result of [15] on the testability of affine-invariant codes with the single local orbit property.

**Theorem 7 ([15])** *If $\mathcal{C} \subseteq \mathbb{F}_2^N$ is linear and has the $k$-single orbit property under the affine group, then $\mathcal{C}$ is $(k, \Omega(1/k^2))$-locally testable.*

We note that in [15] the single-orbit property under the affine group is described as the "strong formal characterization".

Our main theorem, Theorem 4, when combined with the above theorem, immediately yields the following implication for sparse affine invariant codes.

**Corollary 8** *For every constant $t$ there exists a constant $k$ such that if $\mathcal{C} \subseteq \mathbb{F}_2^N$ is a linear, affine-invariant code with at most $N^t$ codewords, then $\mathcal{C}$ is $(k, \Omega(1/k^2))$-locally testable.*

## 2.4 Implications to BCH codes

In addition to the implications for the testability of sparse affine-invariant codes, our results also give new structural insight into the classical BCH codes. Even though these codes have been around a long time, and used often in the CS literature, some very basic questions about them are little understood. We describe the codes, the unanswered questions about them, and the implications of our work in this context below.

We start by defining the BCH codes and the extended-BCH codes. The former are classical cyclic codes, and the latter are affine-invariant.

Let $\text{Trace} : \mathbb{F}_{2^n} \to \mathbb{F}_2$ be the function $\text{Trace}(x) = x + x^2 + \ldots + x^{2^{n-1}}$. We define the BCH codes by defining their dual.

**Definition 9** *For every pair of integers $n$ and $t$, the (binary) dual-BCH code with parameters $n$ and $t$, denoted $\text{BCH}(n,t)^\perp \subseteq \mathbb{F}_2^{N-1}$ consists of the evaluations of traces of polynomials of degree $2t$ over $\mathbb{F}_{2^n}^*$. I.e.,*

$$\text{BCH}(n,t)^\perp = \{\langle \text{Trace}(f(\alpha)) \rangle_{\alpha \in \mathbb{F}_{2^n}^*} | f \in \mathbb{F}_{2^n}[x], \deg(f) \leq 2t\}$$

*The BCH code* $\mathrm{BCH}(n,t)$ *is simply the dual of* $\mathrm{BCH}(n,t)^\perp$.

*The extended dual-BCH code* $\mathrm{eBCH}(n,t)^\perp \subseteq \mathbb{F}_2^N$ *is simply the evaluation of the same functions over all of* $\mathbb{F}_{2^n}$, *and* $\mathrm{eBCH}(n,t)$ *is its dual.*

(We note that the more common definition of BCH codes is as the subfield subcodes of Reed Solomon codes, with $\mathrm{BCH}(n,t)$ being the subfield subcodes of RS codes of degree $N - 2t - 1$. But it is a folklore fact that the two definitions are equivalent.)

Even though the BCH codes are very classical codes, much is unknown about them. For instance, while it is easy to see (by a counting argument) that the BCH code $\mathrm{BCH}(n,t)$ must have codewords of weight $2t + 1$, such words are not known "explicitly". Till recently it was not known that the set of codes of low weight even generate the BCH code, and this was answered affirmatively only recently by Kaufman and Litsyn [12] who showed that words of weight $2t + 1$ and $2t + 2$ certainly include a basis for the BCH code. This proof remains "non-explicit" and the most "succinct" description of this basis is via $O(Nt)$ field elements of $\mathbb{F}_{2^n}$.

Our result manages to make progress on the second question (that of finding an explicit basis) without making progress on the first, by showing that the affine orbit (or in some cases the cyclic orbit) of a single low-weight codeword gives a basis for the BCH code. While this single codeword is still not explicit, the rest of the basis is explicit given the codeword! We state these implications formally below.

**Corollary 10** *For every $t$ there exists a $k$ such that for all prime $n$, $\mathrm{eBCH}(n,t)$ has the $k$-single orbit property under the affine group.*

The above follows from Theorem 4 using the observation that $\mathrm{eBCH}(n,t)^\perp$ is sparse (has $N^{O(t)}$ codewords) and affine invariant.

**Corollary 11** *For every $t$ there exists a $k$ such that for all $n$ such that $2^n - 1$ is prime, $\mathrm{BCH}(n,t)$ has the $k$-single orbit property under the cyclic group.*

The above follows from Theorem 5 using the observation that $\mathrm{BCH}(n,t)^\perp$ is sparse (has $N^{O(t)}$ codewords) and cyclic invariant.

We remark that questions of this nature are relevant not only to coding theory, but also to computing. For instance a recurring question in CS is to find explicit balls of small radius in tightly packed codes that contain many codewords. While we do not make progress toward such questions here, we believe that such questions face difficulty similar to ours. In particular these questions need to find explicit low-weight vectors (not in the code) that contain many low-weight codewords.

Finally, we point out that the need for various parameters ($n$ and $2^n - 1$) being prime is a consequence of the application of some recent results in additive number theory that we use to show that certain codes have very high distance. We do not believe such assumptions ought to be necessary; however we do not see any immediate path to resolving the "stronger" number-theoretic questions that would arise by allowing $n$ to be non-prime.

## 3 Overview of techniques

Our main theorems are proved essentially by implementing the following plan:

1. We first show that every codeword in the codes we consider are expressible as the Traces of *sparse* polynomials. In the affine-invariant case we also show that these polynomials have somewhat low-degree, i.e., at most $N^{1-\epsilon}$. This part follows standard literature in coding theory (and similar steps were employed already in [15]).

2. We then apply the recent results in additive number theory to conclude that these codes have very high distance. This already suffices to show that the affine-invariant codes are testable by [14]. However the tests given there are "non-explicit" and we need to work further to get an "explicit" test for these codes, or to show the single-orbit condition.

3. The final, and the novel part of this work, is to show by a counting argument, that there exists one (in fact many) low-weight codewords in the dual of the codes we consider such that their orbit spans the dual.

We elaborate on these steps in detail below, laying out precise statements we will prove.

We start with some notation. Recall $N = 2^n$ and $n$ is prime.

Also, we view elements $c \in \mathbb{F}_2^N$ as functions $c : \mathbb{F}_N \to \mathbb{F}_2$. Let $\{\mathbb{F}_N \to \mathbb{F}_2\}$ denote the set of all such functions. Similarly we view elements $c \in \mathbb{F}_2^{N-1}$ as functions $\mathbb{F}_N^* \to \mathbb{F}_2$ and let $\{\mathbb{F}_N^* \to \mathbb{F}_2\}$ denote the set of all such functions.

For $d \in \{1, \ldots, N-2\}$, let $\mathrm{orb}(d) = \{d, 2d (\bmod N-1), 4d (\bmod N-1), \ldots, 2^{n-1}d (\bmod N-1)\}$. By the primality of $n$, we have that $|\mathrm{orb}(d)| = n$ for every $d$. Let $\mathrm{min\text{-}orb}(d)$ denote the smallest integer in $\mathrm{orb}(d)$, and let $\mathcal{D} = \{\mathrm{min\text{-}orb}(d) \mid d \in \{1, \ldots, N-2\}\} \cup \{N-1\}$. Note that $|\mathcal{D}| = 1 + (N-2)/n$.

For $D \subseteq \mathcal{D}$ let
$$P_{N,D} = \{\alpha_0 + \sum_{d \in D} \alpha_d x^d \mid \alpha_d \in \mathbb{F}_N, \alpha_0, \alpha_{N-1} \in \{0,1\}\},$$

and $P_{N-1,D} = \{\sum_{d \in D} \alpha_d x^d \mid \alpha_d \in \mathbb{F}_N, \alpha_{N-1} \in \{0,1\}\}.$

The first step in our analysis of codes invariant over the affine group (resp. cyclic group) is that such codes can be associated uniquely with a set $D \subseteq \mathcal{D}$ so that every codeword in our code is the evaluation of the trace of a polynomial from the associated family $P_{N,D}$ over $\mathbb{F}_N$ (resp. $P_{N-1,D}$ over $\mathbb{F}_N^*$).

**Lemma 12** *For every cyclic-invariant code $\mathcal{C} \subseteq \{\mathbb{F}_N^* \to \mathbb{F}_2\}$ there exists a set $D \subseteq \mathcal{D}$ such that $c \in \mathcal{C}$ if and only if there exists a polynomial $p \in P_{N-1,D}$ such that $c(x) = \mathrm{Trace}(p(x))$ for every $x \in \mathbb{F}_N^*$. Furthermore $|D| \leq t$ if $|\mathcal{C}| \leq N^t$.*

*Similarly, for every affine-invariant code $\mathcal{C} \subseteq \{\mathbb{F}_N \to \mathbb{F}_2\}$ of cardinality $N^t$, there exists a set $D \subseteq \mathcal{D}$ such that $c \in \mathcal{C}$ if and only if there exists a polynomial $p \in P_{N,D}$ such that $c(x) = \mathrm{Trace}(p(x))$ for every $x \in \mathbb{F}_N$. Furthermore, if $|\mathcal{C}| \leq N^t$, then $|D| \leq t$ and $D \subseteq \{1, \ldots, N^{1-1/t}\}$.*

Thus in both cases codes are represented by collections of $t$-sparse polynomials. And in the affine-invariant case, these are also somewhat low-degree polynomials. In what follows we use $\mathcal{C}_N(D)$ to denote the code $\{\text{Trace}(p(x))|p \in P_{N,D}\}$ and $\mathcal{C}_{N-1}(D)$ to denote the code $\{\text{Trace}(p(x))|p \in P_{N-1,D}\}$.

We next use a (small variant of a) theorem due to Bourgain [3] to conclude that the codes $\mathcal{C}_N(D)$ and $\mathcal{C}_{N-1}(D)$ have very high distance (under the given conditions on $D$).

**Theorem 13 ([3])** *For every $\epsilon > 0$ and $r < \infty$, there is a $\delta > 0$ such that for every prime $n$ the following holds: Let $N = 2^n$ and $\mathbb{F} = \mathbb{F}_N$ and let $f(x) = \sum_{i=1}^{r} a_i x^{k_i} \in \mathbb{F}[x]$ with $a_i \in \mathbb{F}$, satisfy*

1. $1 \leq k_i \leq N - 1$
2. $(k_i, N-1) < N^{1-\epsilon}$ for every $1 \leq i \leq r$
3. $(k_i - k_j, N-1) < N^{1-\epsilon}$ for every $1 \leq i \neq j \leq r$

$$\left| \sum_{x \in \mathbb{F}} (-1)^{\text{Trace}(f(x))} \right| < N^{1-\delta}.$$

We note that strictly speaking, [3, Theorem 7], only considers the case where $N$ is prime, and considers the sum of any character from $\mathbb{F}$ to the complexes (not just $(-1)^{\text{Trace}(\cdot)}$). We note that the proof extends to cases where $N = 2^n$ where $n$ is prime as well. We comment on the places where the proof in [3] (and related papers) have to be changed to get the result in our case, in Appendix A.

In our language the above theorem implies that codes represented by sparse polynomials of somewhat low-degree have large distance. Furthermore if the polynomials are sparse, and $N-1$ is prime, then also the codes have large distance. We thus get the following implication.

**Lemma 14** *For every $t$ there exists a $\delta$ such that the following holds for every $N = 2^n$ for prime $n$. Let $\mathcal{D} = \mathcal{D}(N)$ and let $D \subseteq \mathcal{D}$ be of size at most $t$. Then the code $\mathcal{C} = \mathcal{C}_N(D)$ satisfies $\frac{1}{2} - N^{-\delta} \leq \delta(\mathcal{C}) \leq \frac{1}{2} + N^{-\delta}$.*

*Similarly for every $t$ there exists a $\delta$ such that the following holds for for every $N = 2^n$ such that $N - 1$ is prime. Let $\mathcal{D} = \mathcal{D}(N)$ and let $D \subseteq \mathcal{D}$ be of size at most $t$. Then the $\mathcal{C} = \mathcal{C}_{N-1}(D)$ satisfies $\frac{1}{2} - N^{-\delta} \leq \delta(\mathcal{C}) \leq \frac{1}{2} + N^{-\delta}$.*

We remark that such use of results from number theory in coding theory is also common. For example, the distance of the sparse dual-BCH codes is inferred by using the "Weil bound" on exponential sums in a similar manner.

We now move to the crucial part of the paper where we attempt to use counting style arguments to claim that the codes we are considering have the single orbit property for small $k$. Here our plan is as follows.

We first use a result from [14] to show that for any specific code $\mathcal{C}$ we consider and for every sufficiently large $k$, its dual has roughly $\binom{N}{k}/|\mathcal{C}|$ codewords of weight $k$ (this bound is tight to within $1 \pm \Theta(1/N^c)$ factor, for large enough $k$ (where $k$ is independent of $N$ and depends only on $t$, $c$ and the $\delta$ of Lemma 14). Specifically they show:

**Theorem 15 ([14] Lemma 3.5)** *For every $c, t < \infty$ and $\delta > 0$ there exists a $k_0$ such that for every $k \geq k_0$ and for every code $\mathcal{C} \subseteq \mathbb{F}_2^N$ with at most $N^t$ codewords satisfying $\frac{1}{2} - N^{-\delta} \leq \delta(\mathcal{C}) \leq \frac{1}{2} + N^{-\delta}$ it is the case the $\mathcal{C}^\perp$ has $\binom{N}{k}/|\mathcal{C}| \cdot (1 \pm \theta(N^{-c}))$ codewords of weight $k$.*

Thus for any code $\mathcal{C} = \mathcal{C}(D)$ under consideration, this allows us to conclude that $\mathcal{C}^\perp$ has many codewords of weight $k$ (for sufficiently large, but constant $k$). What remains to be shown is that the orbit of one of these, under the appropriate group (affine or cyclic) contains a basis for the whole code $\mathcal{C}^\perp$. To do so, we consider any codeword $x$ of weight $k$ in the dual whose orbit under the group does *not* contain a basis for $\mathcal{C}^\perp$ (i.e., $\mathrm{Span}(\{x \circ \pi | \pi\}) \neq \mathcal{C}^\perp$). We show that every such word $x$ there is a set $D' \subseteq \mathcal{D}$ of size $|D'| = |D| + 1$ such that $x \in \mathcal{C}(D')^\perp$. The size of $\mathcal{C}(D')$ is roughly a factor of $N$ larger than the size of $\mathcal{C}$ and thus $\mathcal{C}(D')^\perp$ is smaller than $\mathcal{C}^\perp$ by a factor of roughly $N$. We argue further that this code $\mathcal{C}(D')$ also satisfies the same invariant structure as $\mathcal{C}$ and so one can apply Lemma 14 and Theorem 15 to it and thereby conclude that the number of weight $k$ codewords in $\mathcal{C}(D')^\perp$ are also smaller than the number weight $k$ codewords in $\mathcal{C}^\perp$ by a factor of approximately $N$. Finally we notice that the number of sets $D'$ is $o(N)$ and so the set $\cup_{D'} \mathcal{C}(D')^\perp$ can not include all possible weight $k$ codewords in $\mathcal{C}^\perp$, yielding the $k$-single orbit property for $\mathcal{C}$. This leads to the proofs of Theorem 4 and 5 - see Section 5.

## 4 Representing sparse invariant codes by sparse polynomials

In this section we study representations of affine-invariant and cyclic-invariant codes by polynomials and in particular prove Lemma 12. (We will be using the definitions of the sets $\mathcal{D}$, $P_{N,\mathcal{D}}$, and $P_{N-1,\mathcal{D}}$ as defined in Section 3 heavily throughout this section.)

We start by recalling some standard properties of the Trace function. Recall that $\mathrm{Trace}(x) = x + x^2 + x^4 + \cdots + x^{2^{n-1}}$. The Trace function is linear, i.e. $\mathrm{Trace}(\alpha+\beta) = \mathrm{Trace}(\alpha) + \mathrm{Trace}(\beta)\ \forall \alpha, \beta \in \mathbb{F}_N$. Recall that every function from $\mathbb{F}_N$ to $\mathbb{F}_N$ and hence every function from $\mathbb{F}_N$ to $\mathbb{F}_2$ is the evaluation of polynomial from $\mathbb{F}_N[x]$. More useful to us is the fact that every function from $\mathbb{F}_N$ to $\mathbb{F}_2$ can also be expressed as the trace of a polynomial from $\mathbb{F}_N[x]$, however this representation is not unique. E.g., $\mathrm{Trace}(x^d) = \mathrm{Trace}(x^{2d}) = \mathrm{Trace}(x^{2^i \cdot d})$. However if we restrict to the setting of polynomials from $P_{N,\mathcal{D}}$ then this representation is unique, as shown below.

**Lemma 16** *For every word $w : \mathbb{F}_N \to \mathbb{F}_2$ (respectively $w : \mathbb{F}_N^* \to \mathbb{F}_2$) there is a unique polynomial $p \in P_{N,\mathcal{D}}$ (respectively $p \in P_{N-1,\mathcal{D}}$) such that $w(x) = \mathrm{Trace}(p(x))$.*

**Proof:** Since every function $w : \mathbb{F}_N \to \mathbb{F}_N$, we can write $w(x)$ uniquely as $\sum_{i=0}^{N-1} c_i x^i$ for some coefficients $c_i \in \mathbb{F}_N$. The condition that $w(\alpha) \in \{0, 1\}$ for every $\alpha \in \mathbb{F}_N$, yields some constraints on $c_i$. In particular we have $w(\alpha)^2 = w(\alpha)$ for every $\alpha \in \mathbb{F}_N$ and so $w(x)^2 = w(x) \pmod{x^N - x}$. But $w(x)^2 = \sum_{i=0}^{N-1} c_i^2 x^{2i}$ and so, equating coefficients we have, $c_0^2 = c_0$, $c_{N-1}^2 = c_{N-1}$, and $c_{2i \pmod{N-1}} = c_i^2$ for every $i \in \{1, \ldots, N-2\}$. Thus writing the set $\{0, \ldots, N-1\}$ (the set of degrees of $x$) as $\{0, N-1\} \cup (\cup_{d \in \mathcal{D} - \{N-1\}} \mathrm{orb}(d))$, where the sets $\mathrm{orb}(d)$ are disjoint, we have that $w(x) = c_0 x^0 + c_{N-1} x^{N-1} + \sum_{d \in \mathcal{D} - \{N-1\}} \mathrm{Trace}(c_d x^d)$. Furthermore $c_0, c_{N-1} \in \mathbb{F}_2$ (since $c_0^2 = c_0$ and $c_{N-1}^2 = c_{N-1}$). Finally, using the fact that $\mathrm{Trace}(a) = a$ for $a \in \mathbb{F}_2$ (using the fact that $n$ is odd), we have $w(x) = \mathrm{Trace}(p(x))$ where $p(x) = c_0 x^0 + c_{N-1} x^{N-1} + \sum_{d \in \mathcal{D} - \{N-1\}} c_d x^d$, which is by definition a member of $P_{N,\mathcal{D}}$. This concludes the proof for the case of functions mapping $\mathbb{F}_N$ to $\mathbb{F}_2$. For the case of functions $w : \mathbb{F}_N^* \to \mathbb{F}_2$, the proof is similar except we start by writing $w$ uniquely as $\sum_{i=1}^{N-1} c_i x^i$ (and so $x^{N-1}$ plays the role of the constant function 1). ∎

**Lemma 17** *Suppose $\mathcal{C} \subseteq \{\mathbb{F}_N \to \mathbb{F}_2\}$ is an affine invariant code containing the word $w = \text{Trace}(p(x))$ for some $p \in P_{N,\mathcal{D}}$. Then, for every monomial $x^e$ in the support of $p$, the function $\text{Trace}(x^e)$ is in $\mathcal{C}$. Furthermore, if $e \notin \{0, N-1\}$ then for every $\beta \in \mathbb{F}_N$, $\text{Trace}(\beta x^e) \in \mathcal{C}$.*

*Similarly if $\mathcal{C} \subseteq \{\mathbb{F}_N^* \to \mathbb{F}_2\}$ is cyclic invariant code containing the word $w = \text{Trace}(p(x)$ for $p \in P_{N-1,\mathcal{D}}$. Then, for every monomial $x^e$ in the support of $p$, the function $\text{Trace}(x^e)$ is in $\mathcal{C}$. If $e \neq N-1$ then for every $\beta \in \mathbb{F}_N$, $\text{Trace}(\beta x^e) \in \mathcal{C}$.*

**Proof:** The proof is essentially from [15]. Since their proof is a bit more complex (and considers more general class of functions and non-prime $n$), we include the proof in our setting for completeness.

We start with the cyclic invariant case. Let $p(x) = \sum_{d \in \mathcal{D}} c_d x^d$, where $c_{N-1} \in \{0,1\}$ and let $w(x) = \text{Trace}(p(x))$. Fix $e$ in the support of $p$. We first consider the case $e \neq N-1$. We wish to show that $\text{Trace}(\beta x^e)$ is in $\mathcal{C}$ for every $\beta \in \mathbb{F}_N$. Note that for every $\alpha \in \mathbb{F}_N^*$, $w(\alpha x)$ is in $\mathcal{C}$ (by the cyclic invariance). Furthermore, the function $\sum_{\alpha \in \mathbb{F}_N^*} \text{Trace}(\alpha^{-e}) w(\alpha x)$ is also in $\mathcal{C}$ (by linearity). But as we show below this term is simply $\text{Trace}(c_e x^e)$.

$$\sum_{\alpha \in \mathbb{F}_N^*} \text{Trace}(\alpha^{-e}) w(\alpha x) = \sum_{\alpha \in \mathbb{F}_N^*} \text{Trace}(\alpha^{-e}) \text{Trace}(p(\alpha x))$$

$$= \sum_{\alpha \in \mathbb{F}_N^*} \left( \sum_{j=0}^{n-1} \alpha^{-e \cdot 2^j} \right) \left( \sum_{i=0}^{n-1} \sum_{d \in \mathcal{D}} c_d^{2^i} \alpha^{d \cdot 2^i} x^{d \cdot 2^i} \right)$$

$$= \sum_{j=0}^{n-1} \sum_{i=0}^{n-1} \sum_{d \in \mathcal{D}} c_d^{2^i} x^{d \cdot 2^i} \sum_{\alpha \in \mathbb{F}_N^*} \alpha^{d \cdot 2^i - e \cdot 2^j}$$

Recall that $\sum_{\alpha \in \mathbb{F}_N^*} \alpha^t$ is 0 if $t \neq 0 \pmod{N-1}$ and 1 if $t = 0$. So we conclude that the innermost sum is non-zero only if $d \cdot 2^i = e \cdot 2^j \pmod{N-1}$ which in turn happens only when $d = e$ and $j = i$ (since both $d, e \in \mathcal{D} - \{N-1\}$). We conclude $\sum_{\alpha \in \mathbb{F}_N^*} \text{Trace}(\alpha^{-e}) w(\alpha x) = \sum_{i=0}^{n-1} c_e^{2^i} x^{e \cdot 2^i} = \text{Trace}(c_e x^e)$.

Finally, we need to show that $\text{Trace}(\beta x^e)$ is also in $\mathcal{C}$. To see this, consider the set $S \subseteq \mathbb{F}_N$ defined as $S = \{\gamma | \text{Trace}(c_e \gamma x^e) \in \mathcal{C}\}$. We know $S$ is non-empty (since $1 \in S$), $S$ is closed under addition, and if $\beta \in S$, then so is $\beta \cdot \zeta^e$ for every $\zeta \in \mathbb{F}_N$. Thus, in particular, $S$ contains the set $T = \{p(\omega^e) | p \in \mathbb{F}_2[x]\}$ where $\omega$ is the multiplicative generator of $\mathbb{F}_N^*$. $T$ is again closed under addition and also under multiplication and so is a subfield of $\mathbb{F}_N$. Finally it includes $\omega^e$ as an element and so $T = \mathbb{F}_N$ (the only strict subfield of $\mathbb{F}_N$ is $\mathbb{F}_2$ which does not contain $\omega^e$ for $e \in \mathcal{D}$). We thus conclude that both $S$ and $T$ equal $\mathbb{F}_N$ and so for every $\beta \in \mathbb{F}_N$, $\text{Trace}(\beta x_e) \in \mathcal{C}$.

To prove the lemma for the cyclic invariant case, it remains to consider the case $e = N - 1$. By hypothesis $c_{N-1} = 1$ in this case. Thus we consider the simpler function $\sum_{\alpha \in \mathbb{F}_N^*} w(\alpha x)$ which is also in $\mathcal{C}$. It can be argued as above that this function equals $c_{N-1}x^{N-1} = x^{N-1} = \text{Trace}(x^{N-1})$. This concludes the analysis of the cyclic invariant case.

The affine invariant case is similar (and indeed only needs to use the facts that $w(\alpha x)$ is in $\mathcal{C}$ for every $\alpha \in \mathbb{F}_N$, and the linearity of $\mathcal{C}$). ∎

We now use Lemma 17 to characterize cyclic invariant families, while also working towards the characterization of affine invariant families.

**Lemma 18** *For every affine invariant code $\mathcal{C} \subseteq \{\mathbb{F}_N \to \mathbb{F}_2\}$ there exists a (unique) set $D \subseteq \mathcal{D}$ such that $\mathcal{C} = \{\text{Trace}(p)|p \in P_{N,D}\}$.*

*For every cyclic invariant family $\mathcal{C} \subseteq \{\mathbb{F}_N^* \to \mathbb{F}_2\}$ there exists a (unique) set $D \subseteq \mathcal{D}$ such that $\mathcal{C} = \{\text{Trace}(p)|p \in P_{N-1,D}\}$.*

**Proof:** We start with the affine-invariant case (the cyclic case is almost identical). We let $D$ be the set of all integers $d \in \mathcal{D}$ such that there is some polynomial $p \in P_{N,\mathcal{D}}$ with positive support on the monomial $x^d$ such that $\text{Trace}(p) \in \mathcal{C}$. By Lemma 17 we have that every function $\text{Trace}(\beta x^d) \in \mathcal{C}$ for every $\beta \in \mathbb{F}_N$, if $d \notin \{0, N-1\}$. Furthermore since $\text{Trace}((x+1)^d)$ is also in $\mathcal{C}$, it follows that the constant function 1 is also in $\mathcal{C}$. We conclude that the traces of all the polynomials in $P_{N,D}$ are in $\mathcal{C}$. Conversely, it can also be verified that every function in $\mathcal{C}$ is a trace of a polynomial in $P_{N,D}$.

The cyclic-invariant case is similar. ∎

Lemma 18 essentially suffices to yield Lemma 12 for the cyclic case (though we still need to verify that $|D|$ is small as claimed). For the affine case we need to work a little harder to bound the size of the integers in $D$. To do so we note that affine-invariant properties have further constraints on the set $D$.

For non-negative integers $d$ and $e$ we say $e$ is in the *shadow* of $d$ (denoted $e \prec d$) if in the binary representations $d = \sum_i d_i 2^i$ and $e = \sum_i e_i 2^i$ with $d_i, e_i \in \{0, 1\}$, it is the case that $e_i \leq d_i$ for every $i$. We note that affine-invariant codes are characterized by codes with a "shadow-closure" property described below.

**Lemma 19** *If $\mathcal{C}$ is an affine-invariant code, $\text{Trace}(x^d) \in \mathcal{C}$ and $e \prec d$ then $\text{Trace}(x^e) \in \mathcal{C}$.*

**Proof:** Since $\text{Trace}(x^d) \in \mathcal{C}$ and $\mathcal{C}$ is affine invariant, then $\text{Trace}((x+1)^d) \in \mathcal{C}$. But $(x+1)^d = \prod_i (1+x)^{d_i 2^i} = \prod_i (1 + x^{d_i 2^i}) = \sum_{e \prec d} x^e$. Therefore, $\text{Trace}(\sum_{e \prec d} x^e) \in \mathcal{C}$ and by Lemma 17 $\text{Trace}(x^e) \in \mathcal{C}$. ∎

We can now complete the proof of Lemma 12.

**Proof of Lemma 12.:** For the cyclic invariant case, the lemma is immediate from Lemma 18 which claims that every cyclic invariant code $\mathcal{C} = \mathcal{C}_{N-1}(D) = \{\text{Trace}(p)|p \in P_{N-1,D}\}$ for some $D \subseteq \mathcal{D}$. Conversely, it can be verified that for every $D \subseteq \mathcal{D}$, the code $\mathcal{C}(D)$ is cyclic invariant and maps $\mathbb{F}_N^*$ to $\mathbb{F}_2$. Finally, since for every pair of functions $p_1 \neq p_2 \in P_{N-1,D}$ $\text{Trace}(p_1) \neq \text{Trace}(p_2)$, we have that $|\mathcal{C}| = |P_{N-1,D}| \geq N^{|D|}$ yielding $|D| \leq t$ if $|\mathcal{C}| \leq N^t$.

We now consider the affine invariant case. Consider an affine-invariant code $\mathcal{C}$. By Lemma 18 there is a set $D \subseteq \mathcal{D}$ such that $\mathcal{C} = \mathcal{C}_N(D) = \{\text{Trace}(p) | p \in P_{N,D}\}$. As above we also have $|D| \leq t$ if $|\mathcal{C}| \leq N^t$. It remains to be shown that $D \subseteq \{1, \ldots, N^{1-1/t}\}$.

For this part we use Lemma 19 to note first that the set $D$ should be "shadow-closed", i.e., if $d \in D$ and $e \prec d$ then $e \in D$. Now consider the "binary weight" of the integers $d \in D$, i.e., the number of non-zero bits in the binary representation of $d$. We claim that for every integer $d \in D$, its binary weight is (very crudely) at most $t$ (or else its shadow and hence $D$ has more than $t$ elements). It follows that the integer $d = \text{min-orb}(d) \leq 2^{n(1-1/t)} = N^{1-1/t}$. Since this holds for every $d \in D$, we conclude that $D \subseteq \{1, \ldots, \lfloor N^{1-1/t} \rfloor\}$. This yields the proof of Lemma 12 for the affine-invariant case. ∎

## 5 Proofs of Main theorems

We now derive the proofs of the main theorems.

### 5.1 Analysis of the cyclic case

**Proof of Theorem 5:** Let $\delta = \delta(t)$ and $\delta' = \delta'(t+1)$ be as given by Lemma 14 for the cyclic invariant case (so codes of length $N-1$ have distance roughly $1/2 - N^{-\delta}$). Let $c = 2$ and let $k_0 = k_0(c, t, \delta)$ and $k'_0 = k_0(c, t+1, \delta')$ be as given by Theorem 15. We prove the theorem for $k = \max\{k_0, k'_0\}$.

Fix $N$ so that $N - 1$ is prime and let $\mathcal{C} \subseteq \{\mathbb{F}_N^* \to \mathbb{F}_2\}$ be a cyclic code of cardinality at most $N^t$. Let $D \subseteq \mathcal{D}$ be as given by Lemma 12, so that $\mathcal{C} = \{\text{Trace}(p) | p \in P_{N-1,D}\}$. For $d \in \mathcal{D} - D$, let $\mathcal{C}(d) = \{\text{Trace}(p) | p \in P_{N-1, D \cup \{d\}}\}$. Our analysis below will show that (1) Every codeword in $w \in \mathcal{C}^\perp - \cup_{d \in \mathcal{D}-D}(\mathcal{C}(d)^\perp)$ generates the code $\mathcal{C}^\perp$ by its cyclic shifts, i.e., $\mathcal{C}^\perp = \text{Span}\{w(\alpha x) | \alpha \in \mathbb{F}_N^*\}$, and (2) There is a codeword of weight $k$ in $\mathcal{C}^\perp - \cup_{d \in \mathcal{D}-D}(\mathcal{C}(d)^\perp)$. Putting the two together we get the proof of the theorem.

We start with the first part. Consider any codeword $w \in \mathcal{C}^\perp$. We claim that if $\text{Span}\{w(\alpha x)\} \neq \mathcal{C}^\perp$, then there must exist an element $d \in \mathcal{D} - D$ such that $w \in \mathcal{C}(d)^\perp$. To see this, first note that $\text{Span}\{w(\alpha x)\}$ is a code invariant under the cyclic group, and is contained in $\mathcal{C}^\perp$. Thus if $\text{Span}\{w(\alpha x)\} \neq \mathcal{C}^\perp$ then it must be strictly contained in $\mathcal{C}^\perp$ and so $(\text{Span}\{w(\alpha x)\})^\perp$ must be a strict superset of $\mathcal{C}$. Using Lemma 12 there must exist a set $D'$ such that $(\text{Span}\{w(\alpha x)\})^\perp = P_{N-1, D'}$. Furthermore $D'$ must be a strict superset of $D$ and so there must exist an element $d \in D' - D$. We claim that $w \in \mathcal{C}(d)^\perp$. This is so since $\mathcal{C}(d) \subseteq (\text{Span}\{w(\alpha x)\})^\perp$ and so $w \in (\text{Span}\{w(\alpha x)\}) \subseteq \mathcal{C}(d)^\perp$. This concludes the proof of the first claim.

It remains to show that there is a codeword of weight $k$ in $\mathcal{C}^\perp - \cup_{d \in \mathcal{D}-D}(\mathcal{C}(d)^\perp)$. For this we employ simple counting arguments. We first note that, using Lemma 14, that $\mathcal{C}$ is a code satisfying $\frac{1}{2} - N^{-\delta} \leq \delta(\mathcal{C}) \leq \frac{1}{2} + N^{-\delta}$. Hence we can apply Theorem 15 to conclude that $\mathcal{C}^\perp$ has at least $\binom{N}{k}/(|\mathcal{C}|) \cdot (1 - O(1/N^2))$ codewords of weight $k$. On the other hand, for every fixed $d \in \mathcal{D} - D$, we have (by Lemma 14 again) $\frac{1}{2} - N^{-\delta'} \leq \delta(\mathcal{C}(d)) \leq \frac{1}{2} + N^{-\delta'}$. Again applying Theorem 15 we have $\mathcal{C}(d)^\perp$ has at most $\binom{N}{k}/(|\mathcal{C}(d)|)(1 + O(1/N^2))$ codewords of weight $k$. In case $d = N - 1$, then $|\mathcal{C}(d)| = 2 \cdot |\mathcal{C}|$. In case $d \neq N-1$ then $|\mathcal{C}(d)| = N \cdot |\mathcal{C}|$. Thus we can bound the total number of codewords of weight $k$ in $\cup_{d \in \mathcal{D}-D}\mathcal{C}(d)^\perp$ from above by

$$\binom{N}{k}/(2 \cdot |\mathcal{C}|)(1+O(1/N^2)) + |\mathcal{D}| \times \binom{N}{k}/(N \cdot |\mathcal{C}|)(1+O(1/N^2)) \leq \frac{1}{2|\mathcal{C}|} \cdot \binom{N}{k}(1+1/\log_2 N + O(1/N^2)),$$

where above we use the fact that $|\mathcal{D}| \leq N/\log_2 N$. For sufficiently large $N$ (i.e., when $1/\log_2 N + O(1/N^2) \leq 1/2$) we have that this quantity is strictly smaller than $\binom{N}{k}/(|\mathcal{C}|) \cdot (1 - O(1/N^2))$, which was our lower bound on the number of codewords of weight $k$ in $\mathcal{C}^\perp$. We conclude that there is a codeword of weight $k$ in $\mathcal{C}^\perp - \cup_{d \in \mathcal{D}-D}(\mathcal{C}(d)^\perp)$ as claimed.

This concludes the proof of the theorem. ∎

### 5.2 Analysis of the affine-invariant case

**Proof of Theorem 4:** The proof is similar to the proof of Theorem 5 with the main difference being that we need to argue that the polynomials associated with functions in $\mathcal{C}$ and $\mathcal{C}(d)$ are of somewhat low-degree (to be able to conclude that they have high-distance). Details below.

Given $t$, let $\delta$ be from Lemma 14 and let $k$ be large enough for application of Theorem 15. Fix $N = 2^n$ for prime $n$ and and let $\mathcal{C}$ be an affine-invariant code of cardinality $N^t$. Let $D \subseteq \mathcal{D}$ be a set of cardinality at most $t$ and consisting of integers smaller that $N^{1-1/t}$ such that $\mathcal{C} = \{\text{Trace}(p)|p \in P_{N,D}\}$ (as given by Lemma 12). For $d \in \mathcal{D} - D$, let $\mathcal{C}(d) = \{\text{Trace}(p)|p \in P_{N,D \cup \{d\}}\}$.

Let $\mathcal{D}' = (\mathcal{D} - D) \cap \{1, \ldots, \lfloor N^{1-1/t} \rfloor\}$.

Similar to the proof of Theorem 5 we argue that if there is a weight $k$ codeword $w$ in $\mathcal{C}^\perp$ that is not in some $\mathcal{C}(d)^\perp$, but now only for every $d \in \mathcal{D}'$, then $\{\text{Span}(w(\alpha x + \beta)|\alpha \in \mathbb{F}_N^*, \beta \in \mathbb{F}_N\} = \mathcal{C}^\perp$. The same counting argument as in the proof of Theorem 5 suffices to show that such a word does exist.

Consider $w \in \mathcal{C}^\perp$ and the code $\{\text{Span}(w(\alpha x + \beta)|\alpha \in \mathbb{F}_N^*, \beta \in \mathbb{F}_N\}$. $\{\text{Span}(w(\alpha x + \beta))\}$ is affine invariant and so is given by $P_{N,E}$ for some shadow-closed set $E$. If $\{\text{Span}(w(\alpha x + \beta))\}^\perp \neq \mathcal{C}$ then $E$ strictly contains $D$ and so there must exist some element $d' \in E - D$. Now consider smallest binary weight element $d \prec d'$ such that $d \in E - D$. We claim that the binary weight of $d$ must be at most $t + 1$ (since elements of $D$ have binary weight at most $t$). We then conclude that $w \in \{\text{Span}(w(\alpha x + \beta))\} \subseteq \mathcal{C}(d)^\perp$ yielding the claim.

The counting argument to show there is a codeword of weight $k$ in $\mathcal{C}^\perp - (\cup_{d \in \mathcal{D}'}\mathcal{C}(d)^\perp$ is now same as in the proof of Theorem 5 except that we use the affine-invariant part of Lemma 14.

This completes the proof of Theorem 4. ∎

## Acknowledgments

We would like to thank Oded Goldreich for valuable suggestions and anonymous reviewers for detecting several omissions and errors in a prior version of this paper. We thank Swastik Kopparty for helpful discussions.

## A  On using results from additive number theory

As pointed out earlier Theorem 7 of [3] only considers the analog of Theorem 13 where the field $\mathbb{F}$ is of prime cardinality $N$, and shows that for any additive character $\chi$, $|\sum_{x \in \mathbb{F}} \chi(f(x))| \leq N^{1-\delta}$. Here we mention the modifications necessary to extend the proof to the case where $\mathbb{F}_N$ is of cardinality $2^n$ with $n$ being prime.

In [3] the proof reduces to the two cases $r = 1$ and $r = 2$. The case $r = 1$ in the prime case was obtained in [7]. In our case, where $N = 2^n$, the $r = 1$ case was shown in [6]. For $r = 2$ the proof in the prime case applied the sum-product theorem from [5] and uses Proposition 1 of [4]. We note that Proposition 1 of [4] works also when the field is not of prime cardinality. As argued in [5], the sum-product statement might weaken for more general fields only when the field $\mathbb{F}_N$ contains somewhat large subfields. However, when $n$ is prime $\mathbb{F}_{2^n}$ contains only the constant size base field $\mathbb{F}_2$. We conclude that when $\mathbb{F} = \mathbb{F}_{2^n}$ ($n$ prime) it remains true that if a set $A \subset \mathbb{F}_N$ has size $1 < |A| < N^{1-\epsilon}$ for some given $\epsilon$ then $|A + A| + |A \cdot A| > C|A|^{1+\delta}$, for some $\delta = \delta(\epsilon)$. The key ingredient of the proof in [4] is an additional sum-product theorem in the additive/multiplicative group $\mathbb{F}_N \times \mathbb{F}_N$ with $N$ prime, where addition and multiplication are defined coordinate-wise. The equivalent formulation for our case $\mathbb{F}_{2^n} \times \mathbb{F}_{2^n}$ follows exactly as in [4], and so does the rest of the proof.